\documentclass{PoS}

\title{Exotic and higher spin mesons in charmonium}

\ShortTitle{Exotic and higher spin mesons in the charmonium}

\author{Jo Dudek, Robert Edwards, 
\speaker{Nilmani Mathur}\thanks{Email: nilmani@jlab.org}, and David Richards\\
Thomas Jefferson National Accelerator Facility,
12000 Jefferson Avenue, Newport News, VA 23606, USA}

\abstract{Exotic and higher spin (> 1) mesons are still not throughly
investigated in lattice QCD.  Using a set of derivative based
operators we report our exploratory study of these mesons in
charmonium region.  We use a $12^3 \times 48$ anisotropic ($\xi$ = 3)
clover lattice with inverse temporal lattice spacing $a^{-1}$ =  6.05 GeV.  Techniques
developed in this exploratory study will be utilized in our future
comprehensive study of light hybrid mesons that are to be explored in
the 12 GeV GlueX experiment at Jefferson Laboratory.}

\FullConference{XXIVth International Symposium on Lattice Field Theory\\
                July 23-28, 2006\\
                Tucson, Arizona, USA}

\begin{document}

\newcommand{\half}{\frac{1}{2}}
\newcommand{\be}{\begin{displaymath}}
\newcommand{\ee}{\end{displaymath}}
\newcommand{\bea}{\begin{eqnarray}}
\newcommand{\eea}{\end{eqnarray}}
\newcommand{\bdm}{\begin{displaymath}}
\newcommand{\edm}{\end{displaymath}}
\newcommand{\<}{\langle}
\renewcommand{\>}{\rangle}
\newcommand{\Tr}{\mbox{Tr}}
\section{Introduction}
Recently there has been a considerable resurgence of interest in
charmonium physics, motivated by the observation of a number of
puzzling resonances at $B$ factories, and by the advent of CLEO-c as
well as by the BES upgrade. The QCD spectrum is potentially very rich,
admitting states with high angular momenta, glueballs, four-quark
states, and hybrid mesons with manifest gluonic degrees of freedom,
and therefore numerous ideas have been advanced~\cite{charm_review} to interpret
these newly observed states. To resolve these ideas, a comprehensive,
first-principles calculation of the charmonium spectrum using lattice
QCD is therefore crucial.  Moreover, a detailed lattice study of
hybrid mesons in the charmonium region will help us to develop techniques
that can be used in the light-quark sector, whose exploration is the
goal of the GlueX experiment as part of the $12~{\rm GeV}$ upgrade at
Jefferson Laboratory.

To date, the lattice investigations of the spectrum of exotics and
states with higher angular momentum are quite limited. In the
light-quark sector, though several calculations have been performed to study
the exotic $1^{-+}$~\cite{exotic_ref}, a firm conclusion from lattice QCD about the existence of this state at low pion mass region is yet to be drawn. In the
heavy quark sector only a handful of calculations have been performed
to study hybrids and higher spin states~\cite{manke,
charm_hybrids}. The main difficulty encountered when studying these
states is the poor overlap of these states with the chosen
interpolating operators. Since it is not possible to construct an
exotic-meson interpolating field from local quark bilinears alone, one needs
to include link variables  
directly in the
interpolating field, introducing additional statistical noise.
Similarly, states with higher angular momentum require operators
constructed from displaced fields, likewise introducing statistical
noise.  To efficiently tackle such states requires the use of 
appropriate smearing
of the quark and gluon fields, and a large statistical ensemble.

Besides the spectrum, it is only recently that there has
been a lattice study of transition form factors between charmonium
states~\cite{charm0_jlab}; in two exploratory studies it has been
demonstrated that radiative decays~\cite{charm0_jlab} as well as
two-photon-decay form factors~\cite{2photon_jo} can be studied on the
lattice.  We intend to extend those studies from form factors of
conventional, low-lying charmonium states, to hybrids and other
excited states. A subsequent study in the light quark sector will
provide much needed information on the photocouplings to the light
hybrid mesons relevant for the GlueX experiment.

In these proceedings we report the preliminary results of a
comprehensive study of exotics and higher spin mesons in the charmonium
region, in the quenched approximation to QCD.  We employ as a basis of
operators those introduced in Ref.~\cite{manke}, together with
additional operators missing in their construction.  In forthcoming
publications, we will report in more detail the results on the masses
of ground and excited states, as well as on transition form factors
between various states, including hybrids. These results in the charm
sector will facilitate our similar study in the light-quark sector,
using dynamical fermions.
\vspace*{-0.1in}
\section{Numerical details}
\vspace*{-0.1in}
\subsection{Anisotropic clover action}
The relatively large value of the charm-quark mass, $m_c \sim 1.1 - 1.5~{\rm GeV}$, poses challenges to lattice QCD at the currently
employed inverse lattice spacings of $a^{-1} \simeq 2 - 3~{\rm GeV}$,
since $m_c a $ is then large introducing substantial discretization
uncertainties.  Whilst a non-relativistic action might be appropriate
for the $b$ quark, in the charmonium sector it is more accurate to use a
fully relativistic formulation. 
We employ an anisotropic lattice, with finer lattice spacing in the temporal 
than spatial directions, using a Wilson gauge action and clover fermion
action.  The use of an anisotropic lattice introduces additional
parameters beyond those of the isotropic formulation, which must be
tuned to yield the required anisotropy.  We will now outline this
tuning.

In the quenched approximation to QCD, the anisotropy in the gauge
sector can be tuned before, and independently of, the anisotropy in
the fermion sector.  We follow the approach of Ref.~\cite{Klassen}, 
tuning the action to yield a renormalized
anisotropy $\xi \equiv a/a_0 = 3$, where $a$ and $a_0$ are the
spatial and temporal lattice spacings respectively.

We write the fermion action in the form $\sum \bar \psi Q \psi$, where
\vspace*{-0.05in}
\bea
Q &=& m_0 + \nu_0 \nabla_0 \gamma_0 - \half r_0 a_0 \Delta_0
 + \sum_k \left( \nu \nabla_k \gamma_k- \half r a \Delta_k \right) \nonumber \\
&& - \frac{a}{2} \Big[ \omega_0 \sum_k \sigma_{0k} F_{0k}
 + \omega \sum_{k<l} \sigma_{kl} F_{kl} \Big ] ~,
\label{eq:Q_generic}
\eea 
Detail about this action and its implementation is given in Ref.~\cite{Klassen}.
Classical values of action parameters are given by
~\cite{manke}
\begin{eqnarray}
\vspace*{-0.5in}
m_{0} &=& m_{q} (1+ {1\over 2} m_{s} m_{q})~,\\
\nu_{t} &=& \nu_{s} {{1 + {1\over 2} a_{s} m_{q}}\over{1 + {1\over 2} a_{s} m_{q}}}~,\\
c_{s} &=& \nu_{s}~,\\
c_{t} &=& {1\over 2}{\left(\nu_{s} + \nu_{t} {a_{t}\over{a_{s}}}\right)}~.
\end{eqnarray}
Following Ref.~\cite{manke}, classical values of the clover
coefficients, Eqs.~2.4 and 2.5, are improved by tadpole factors, $u_s$
and $u_t$, for spatial and temporal links, respectively, to yield
\vspace*{-0.1in}
\begin{eqnarray}
c_{s} &\longrightarrow& c_{s}/u_s^{3}~,\\
c_{t} &\longrightarrow& c_{t}/(u_t u_s^{2})~.
\end{eqnarray}
The charm mass is determined by tuning the bare quark mass $m_0$
non-perturbatively such that the spin average of the lowest S-wave
mesons (1S) coincides with its experimental value, i.e. : 
$
(3 m_{J/\Psi} +
 m_{\eta_c})/4 = 3.067~{\rm GeV}.
$
The parameters $\nu_t$ and $\nu_s$ are not independent; we choose
$\nu_t = 1$ and then tune $\nu_s$ non-perturbatively to satisfy the
lattice dispersion relation
\begin{eqnarray}
 c(p)^{2} = {{E(p)^{2} - E(0)^{2}} \over p^{2}} = \xi^{2} {{a_{0}^{2} E(p)^{2} - a_{0}^{2}E(0)^{2}}\over a^{2}p^{2}} = 1.
\end{eqnarray}
Keeping all other parameters fixed, we tune $\nu_s$ to satisfy the
above relation to within $\sim 1\%$.

\subsection{Interpolating operators for exotic and higher spin mesons}
Following Ref.~\cite{manke}, we employ a set of operators for exotic and
higher spin mesons using covariant derivatives on the quark fields. In
Ref.~\cite{manke}, such operators have forward directional
derivatives, $\vec{\nabla}$, i.e., acting only on the quark field. For states at rest, this can create
required quantum numbers. However, for non-zero momentum, the forward
directional derivative alone fails to yield operators of definite
charge conjugation. One needs a backward-forward derivative
${\stackrel {\longleftrightarrow}{\nabla}}$ for this purpose.  Hence
we change the operator table in Ref.~\cite{manke} by replacing
$\vec{\nabla}\rightarrow {\stackrel {\longleftrightarrow}
{\nabla}} = {\stackrel {\longrightarrow} {\nabla}} - {\stackrel {\longleftarrow}
{\nabla}}$. Backward-forward derivatives were also utilized to study
exotics in Ref.~\cite{bing-an}.

We use gauge-invariant quark smearing~\cite{charm0_jlab}, along with
stout-link smearing~\cite{stout}, to obtain better overlap of the
operators to the ground states. Smeared-source point-sink (SP), as
well as smeared-source smeared-sink (SS), correlators are used to
obtain better signal.  Smearing parameters are tuned separately for
the various channels. We observe that displacement lengths of one lattice site
and
two lattice sites yield largely indistinguishable results in effective masses of
these operators, and therefore for this calculation we choose
displacement length one.

\vspace*{-0.1in}
\section{Results}
\vspace*{-0.05in}
We employ 1996 quenched configurations on a $12^3 \times 48$ lattice
(with inverse temporal lattice spacing $6.05~{\rm GeV}^{-1}$, obtained from the
static quark-antiquark potential. To generate meson correlators we used Dirichlet boundary condition with source at 5.

\begin{figure}[ht]
\vspace*{-0.08in}
  \begin{center}
    \includegraphics*[width=7cm, height=4.4cm]{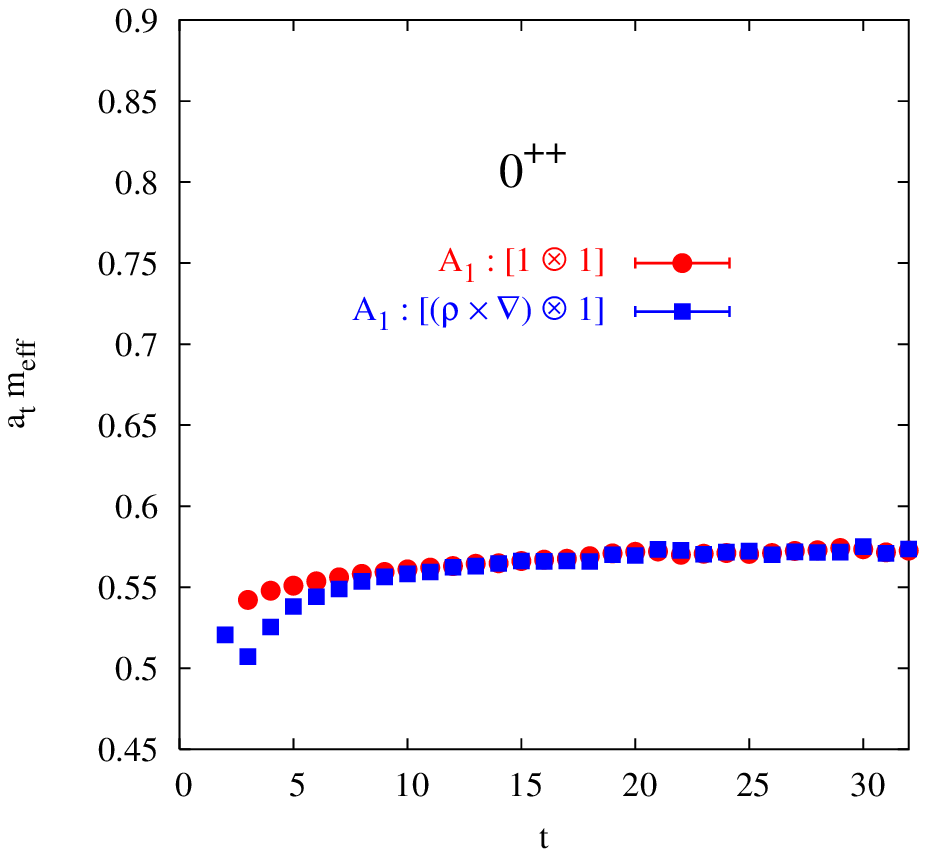}
    \includegraphics*[width=7cm, height=4.4cm]{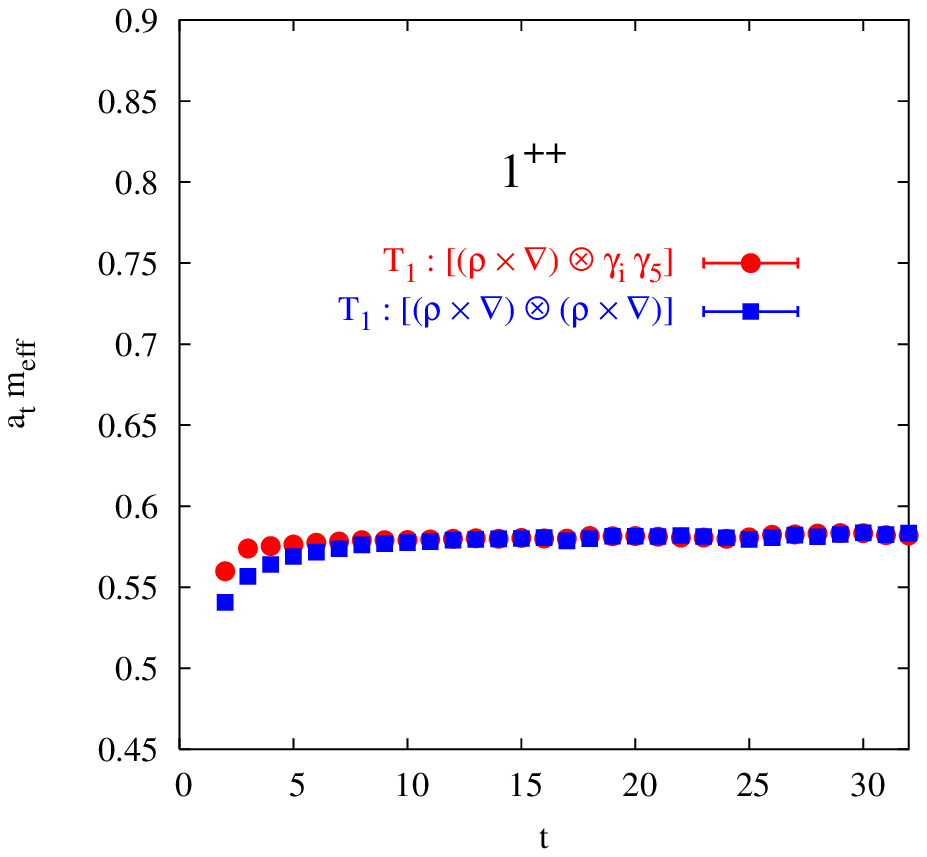}\\
    \includegraphics*[width=7cm, height=4.4cm]{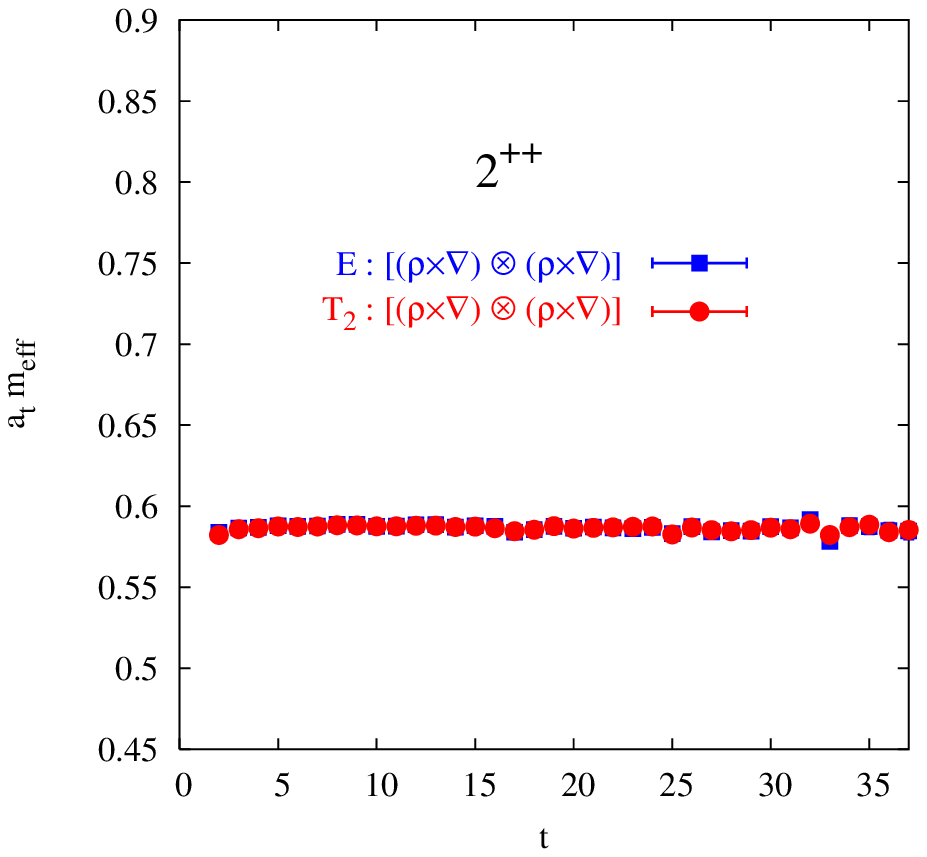}
    \includegraphics*[width=7cm, height=4.4cm]{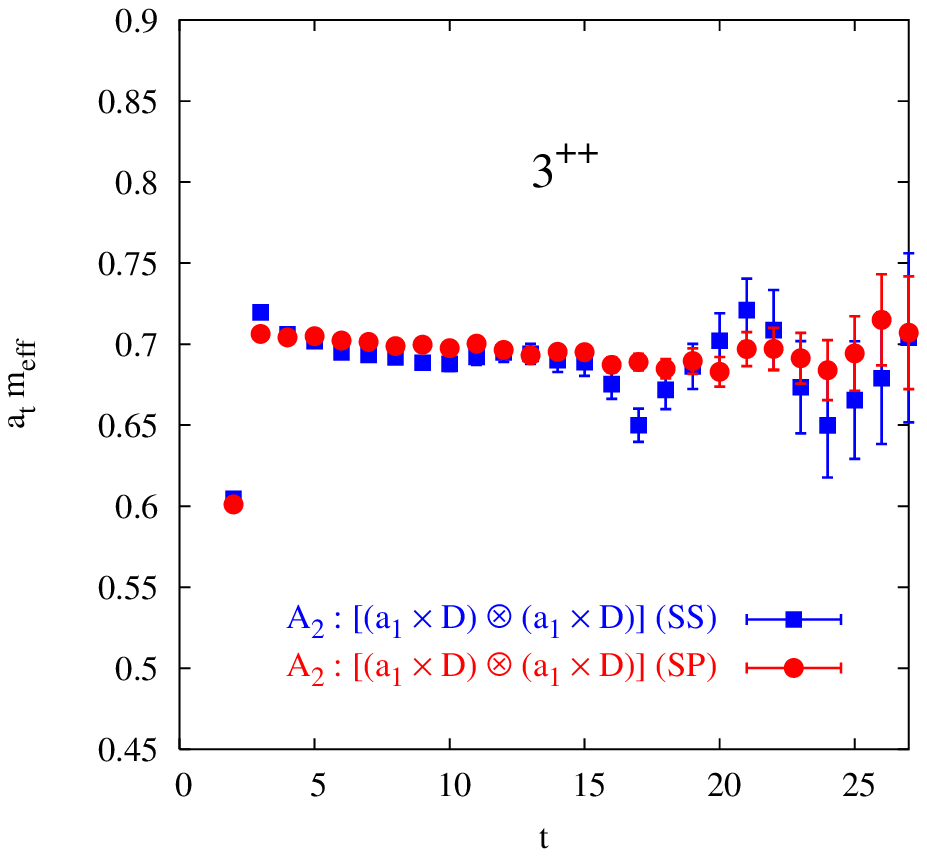}
  \end{center}
\vspace*{-0.3in}
  \caption{Effective masses for $J^{++}$ states with various local and displaced operators.}
    \label{fig:a1}
\end{figure}

\begin{figure}
\vspace*{-0.1in}
  \begin{center}
    \includegraphics*[width=7cm, height=4.4cm]{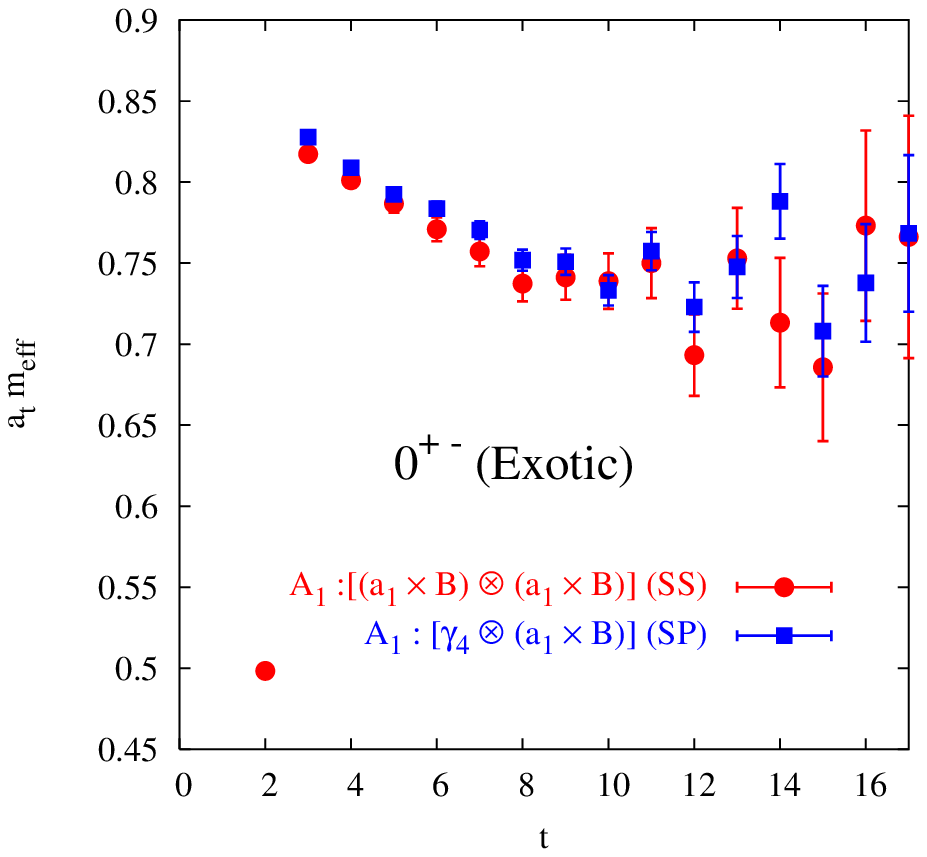}
    \includegraphics*[width=7cm, height=4.4cm]{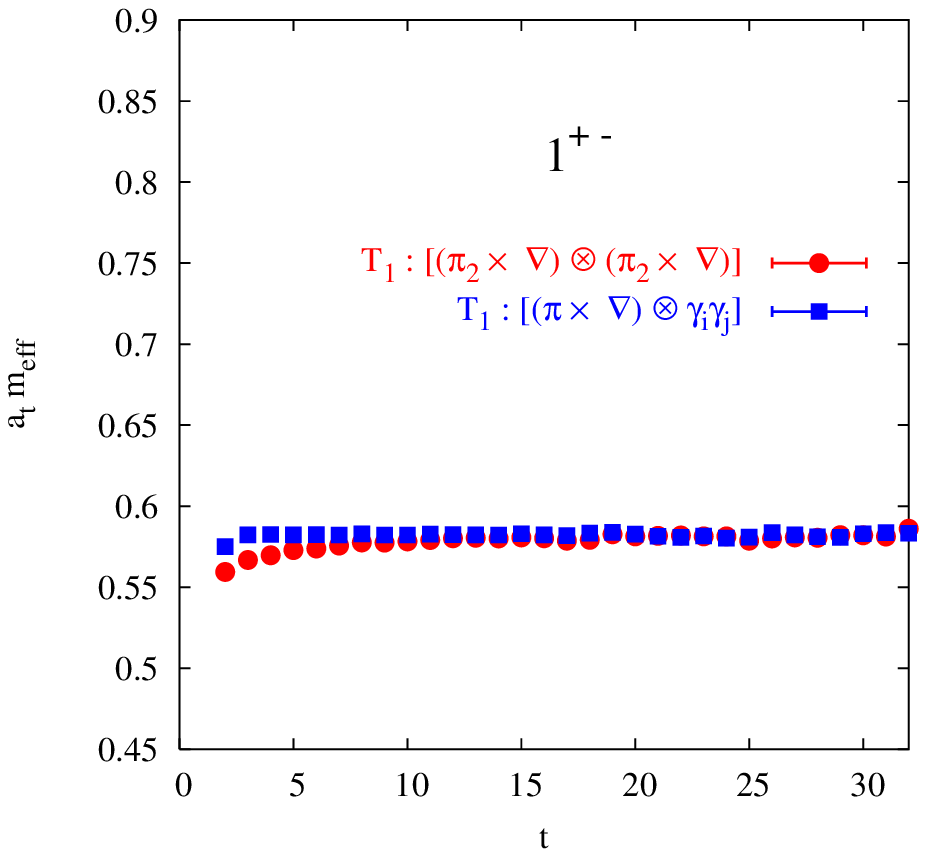}\\
    \includegraphics*[width=7cm, height=4.4cm]{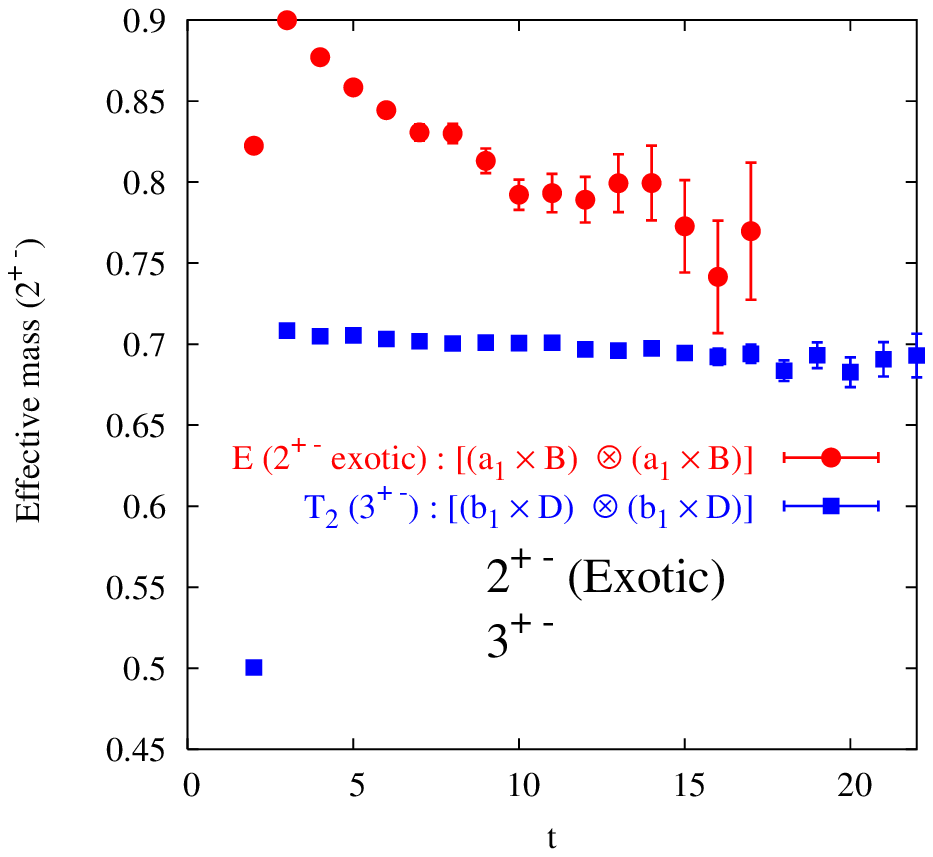}
    \includegraphics*[width=7cm, height=4.4cm]{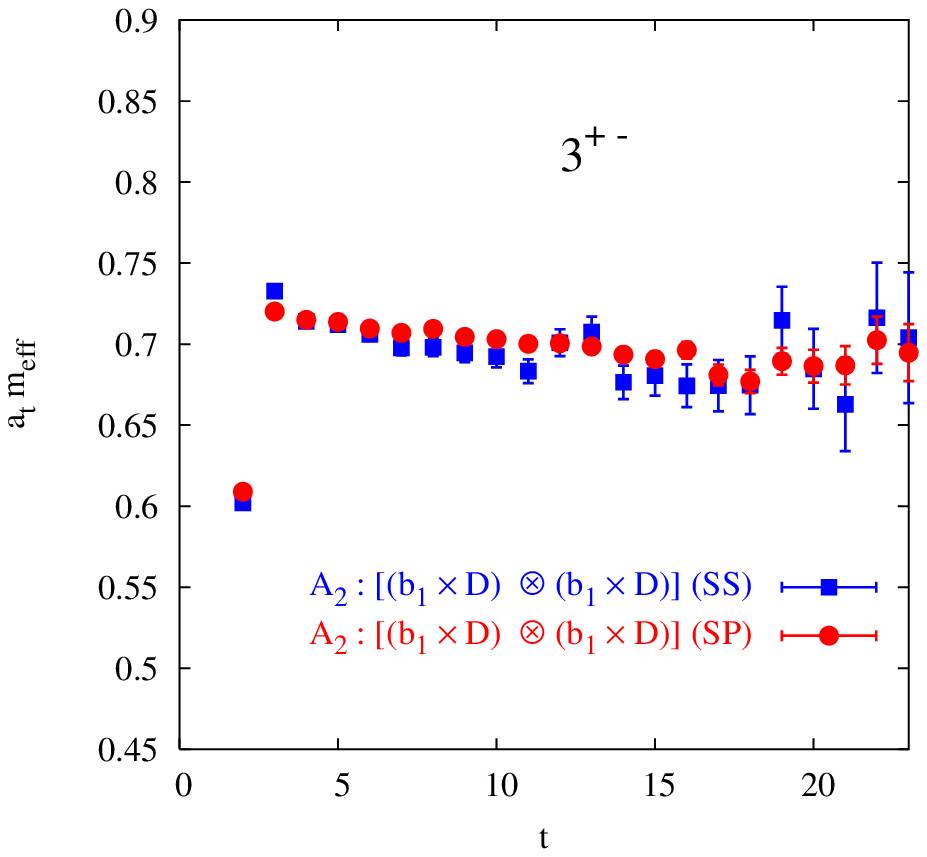}
  \end{center}
\vspace*{-0.3in}
  \caption{Effective masses for $J^{+-}$ states.}
    \label{fig:b1}
\vspace*{-0.2in}
\end{figure}

\begin{figure}
  \begin{center}
\hspace*{0.14in}
    \includegraphics*[width=7cm, height=4.5cm]{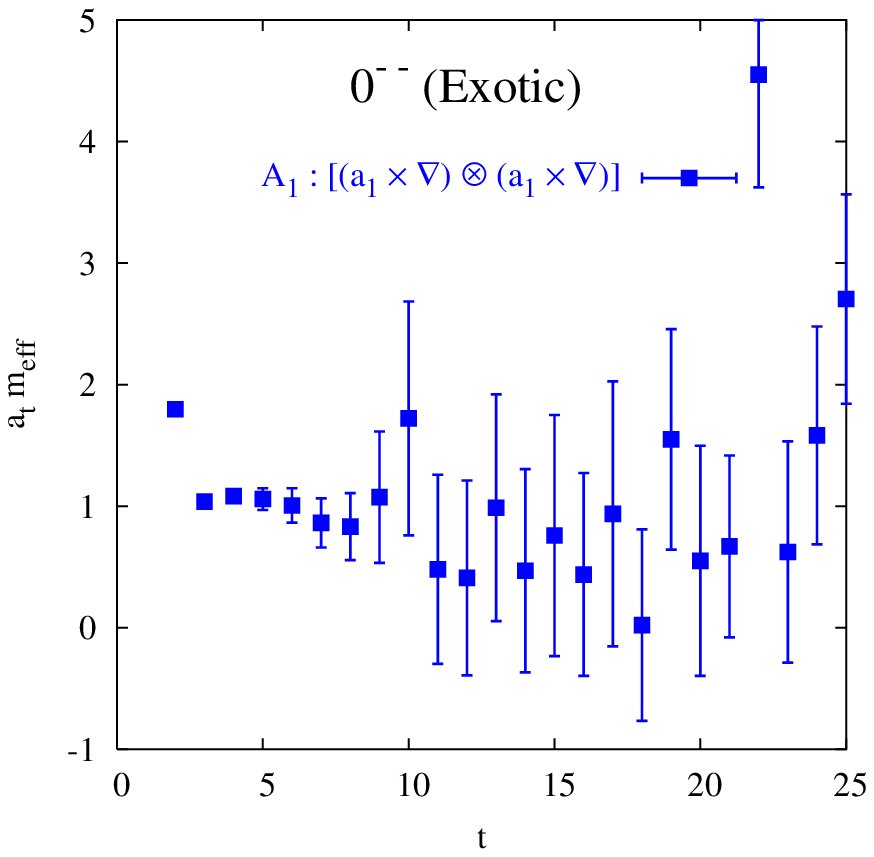}
\hspace*{-0.24in}
    \includegraphics*[width=7cm, height=4.5cm]{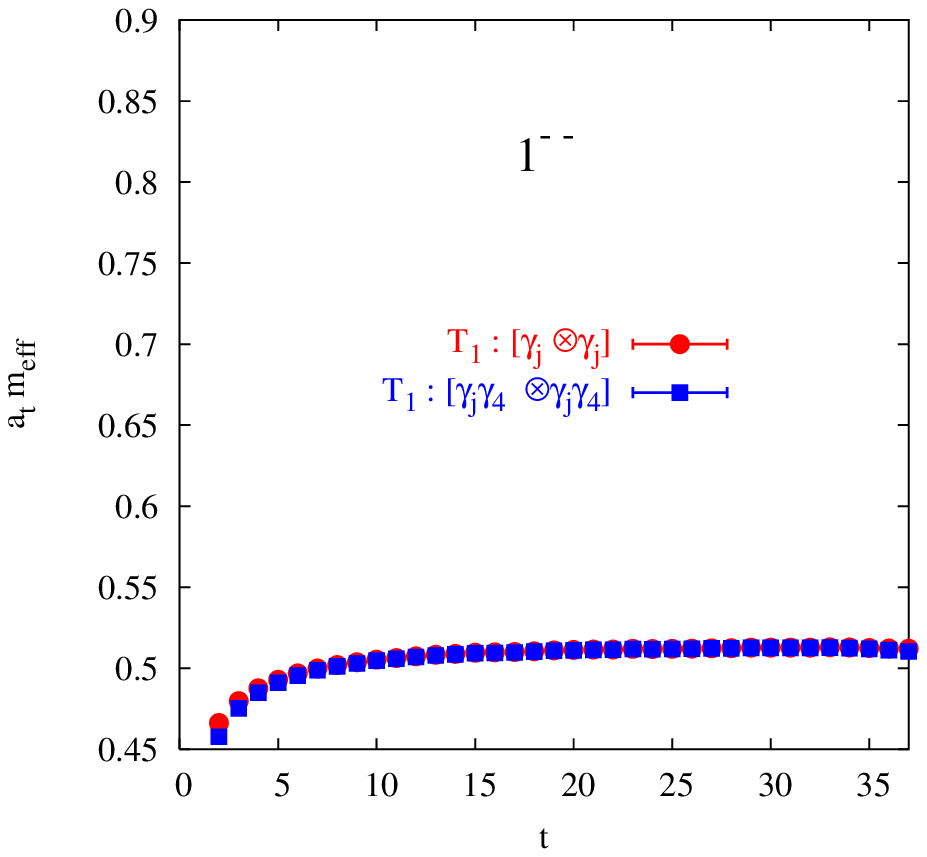}\\
    \includegraphics*[width=7cm, height=4.5cm]{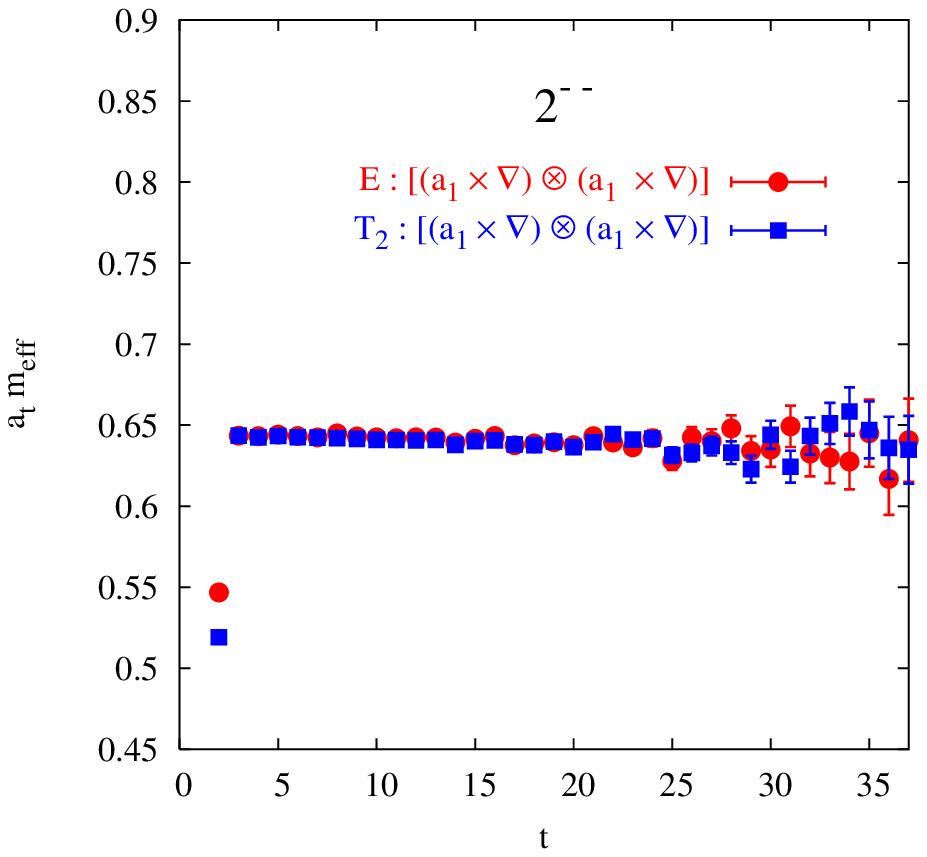}
    \includegraphics*[width=7cm, height=4.5cm]{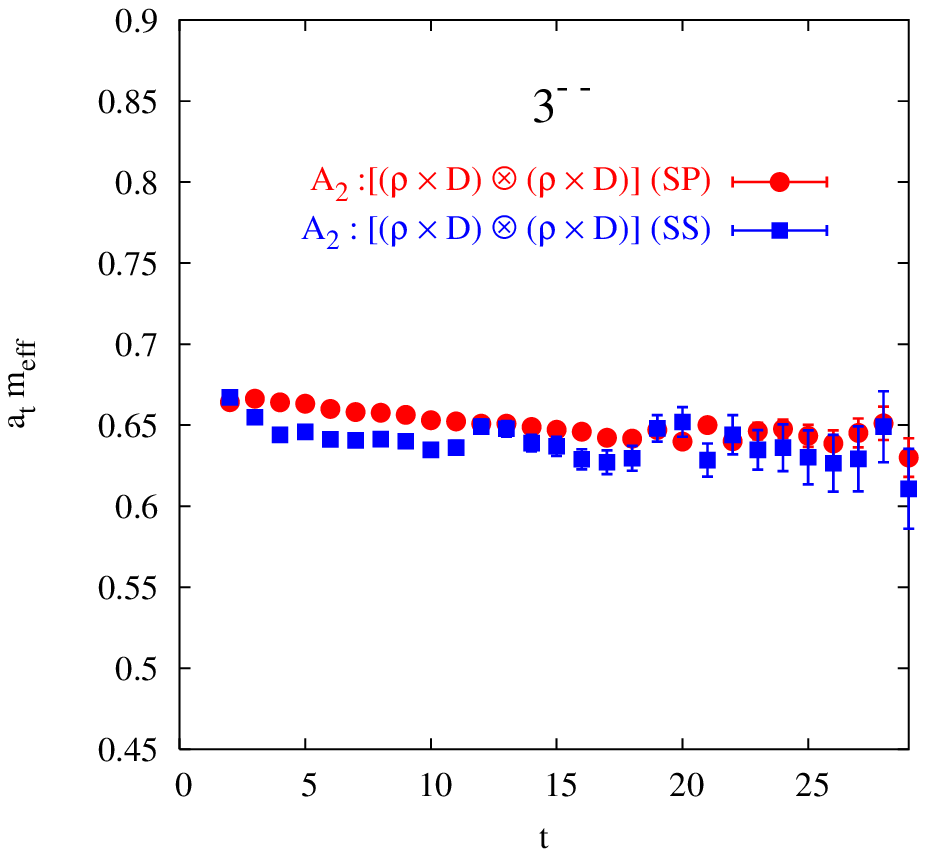}
  \end{center}
  \caption{Effective masses for $J^{--}$ states.}
    \label{fig:rho}
\end{figure}

\begin{figure}
  \begin{center}
    \includegraphics*[width=7cm, height=4.5cm]{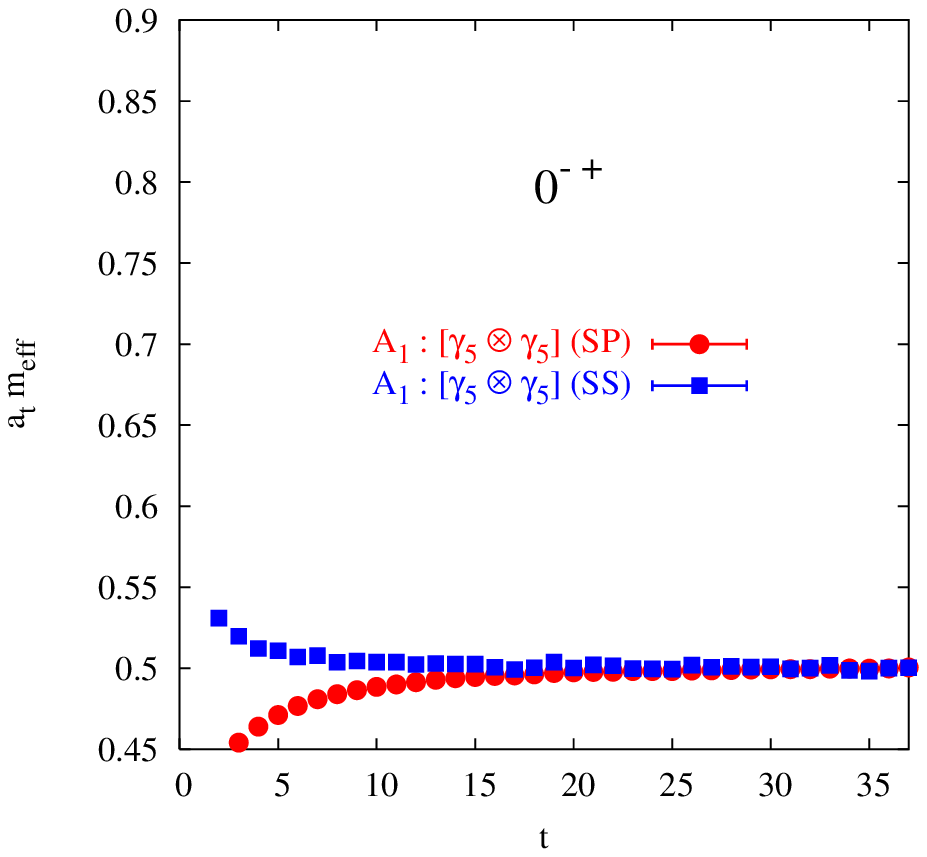}
    \includegraphics*[width=7cm, height=4.5cm]{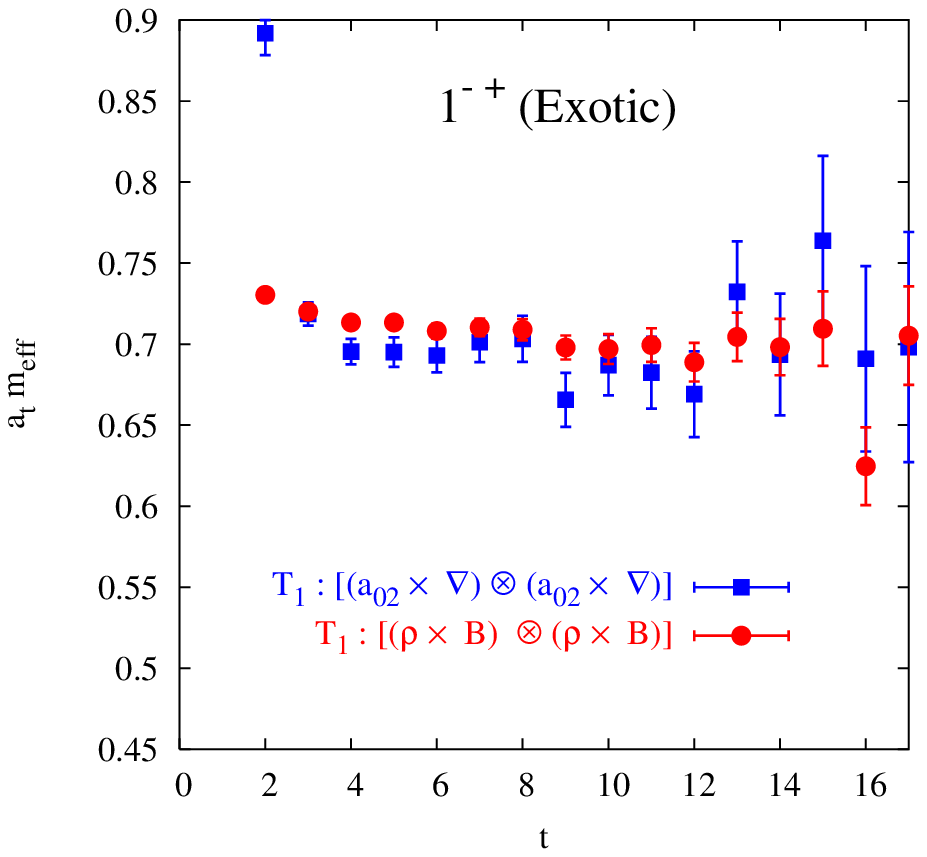}\\
    \includegraphics*[width=7cm, height=4.5cm]{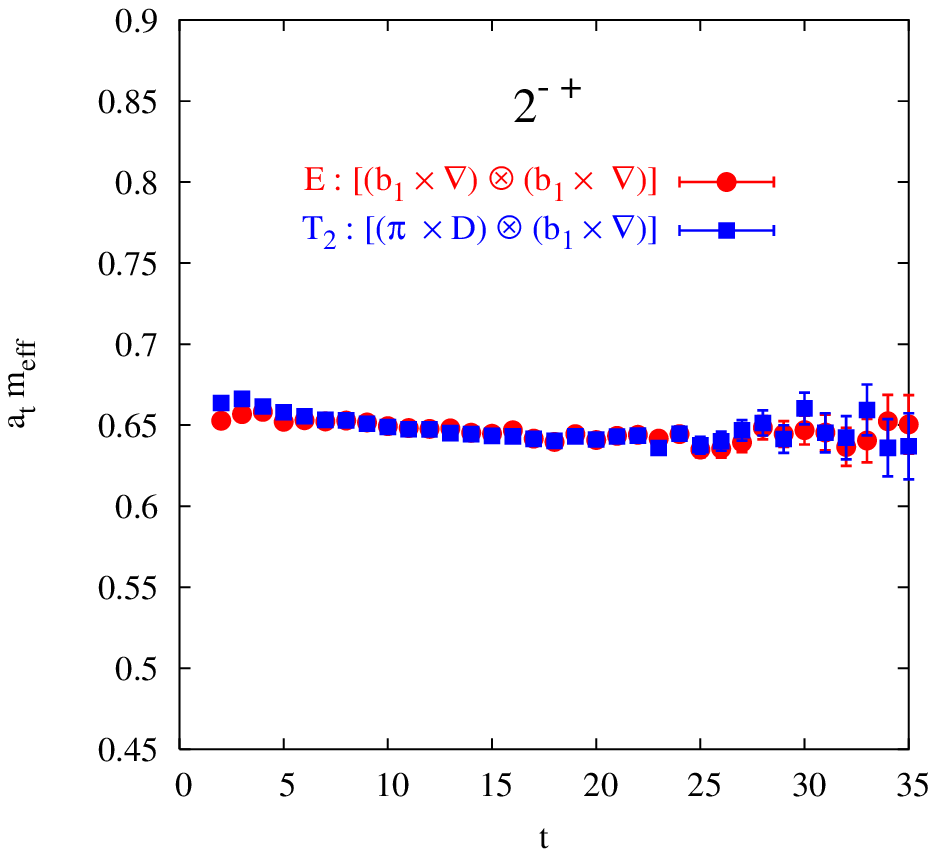}
    \includegraphics*[width=7cm, height=4.5cm]{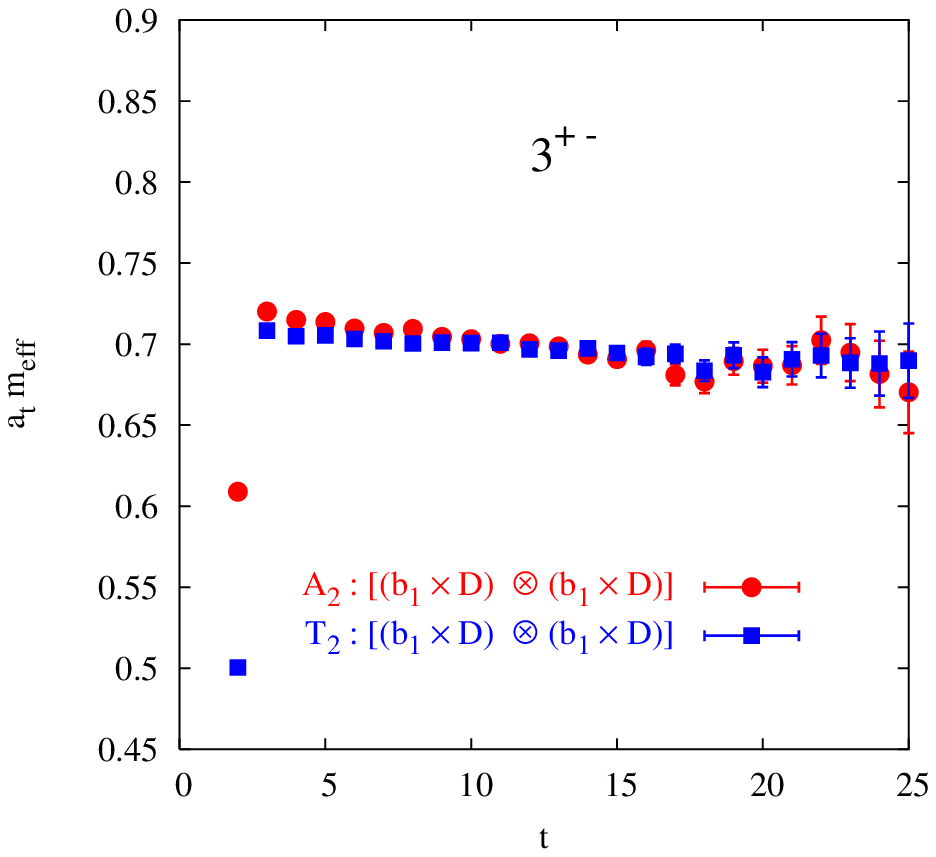}
\vspace*{-0.2in}
  \end{center}
  \caption{Effective masses for $J^{-+}$ states. Bottom right figure is a comparison of  $3^{+-}$ obtained from $A_2$ and $T_2$ representations.}
    \label{fig:pi}
\end{figure}

In Figs.~\ref{fig:a1} - \ref{fig:pi}, we present representative
effective masses for the four different combinations of $PC$, for
various lattice irreducible representations $A_1 (0)$, $A_2(3)$,
$E(2)$, $T_1(1)$ and $T_2(2)$; the numbers in brackets denote the
lowest continuum spin lying in each irrep..  To facilitate a
comparison of the quality of effective masses, we employ the same
scale in each figure, except for the case $A_1^{--} (0^{--})$ where
the signal is noisy and the mass larger. In each channel, including exotics but with the exception of
the $A_1^{--}$, we obtain a very good signal for several
operators. Any of these operators would yield a good ground-state mass
with reasonable statistical errors.  The quality of the data is better
than that of Ref.~\cite{manke},  since we adopted tuned smeared operator
 in part to reduce noise. 
 A detailed
analysis of this data using the variational method, employing
multi-exponential fits, and using Bayesian statistics to yield ground
and excited-state masses will be reported in forthcoming publications.

Even our preliminary analysis enables us to make two observations.
Firstly, the mass of the exotic $1^{-+}$ observed in this study is
lower than that found in Ref.~\cite{manke}, where the effective mass
is poorer {\footnote{although the operators $(\rho \times B)_{T_1}$ and $(a_0 \times \nabla)_{T_1}$ can, at finite a, overlap with $4^{- +}$ (a non-exotic), as well as with $1^{- +}$, in the continuum no such  $4^{- +}$ can remain. As such, we are confident in assigning a $1^{- +}$ status to the observed ground state.}}.  
We also observe a good signal for the other exotics, $0^{+
-}$ and $2^{+ -}$. Secondly, we observe in the $T_2^{+ -}$
channel, the bottom left in Fig.~2, that one operator in this
channel produces a lower effective mass than the others. We believe
this is an overlap to the $3^{+ -}$ state, as $T_2^{+ -}$ operators
have an overlap to both the $2^{+-}$ and $3^{+ -}$ states {\footnote{An operator of the type $(b_{1} \times D)_{T_2}$ retains an overlap with $3^{+ -}$ in the continuum.}}. In the bottom
right figure in Fig.~4, we plot effective masses from the $T_2^{+ -}$
and $A_2^{+ -}$ correlators simultaneously, and observe that they
coincide within statistics. Finally, Fig.~5 shows the relative
position of the various channels, together with the observed
experimental ordering.
\begin{figure}[ht]
  \begin{center}
    \includegraphics*[width=16cm, height=8cm]{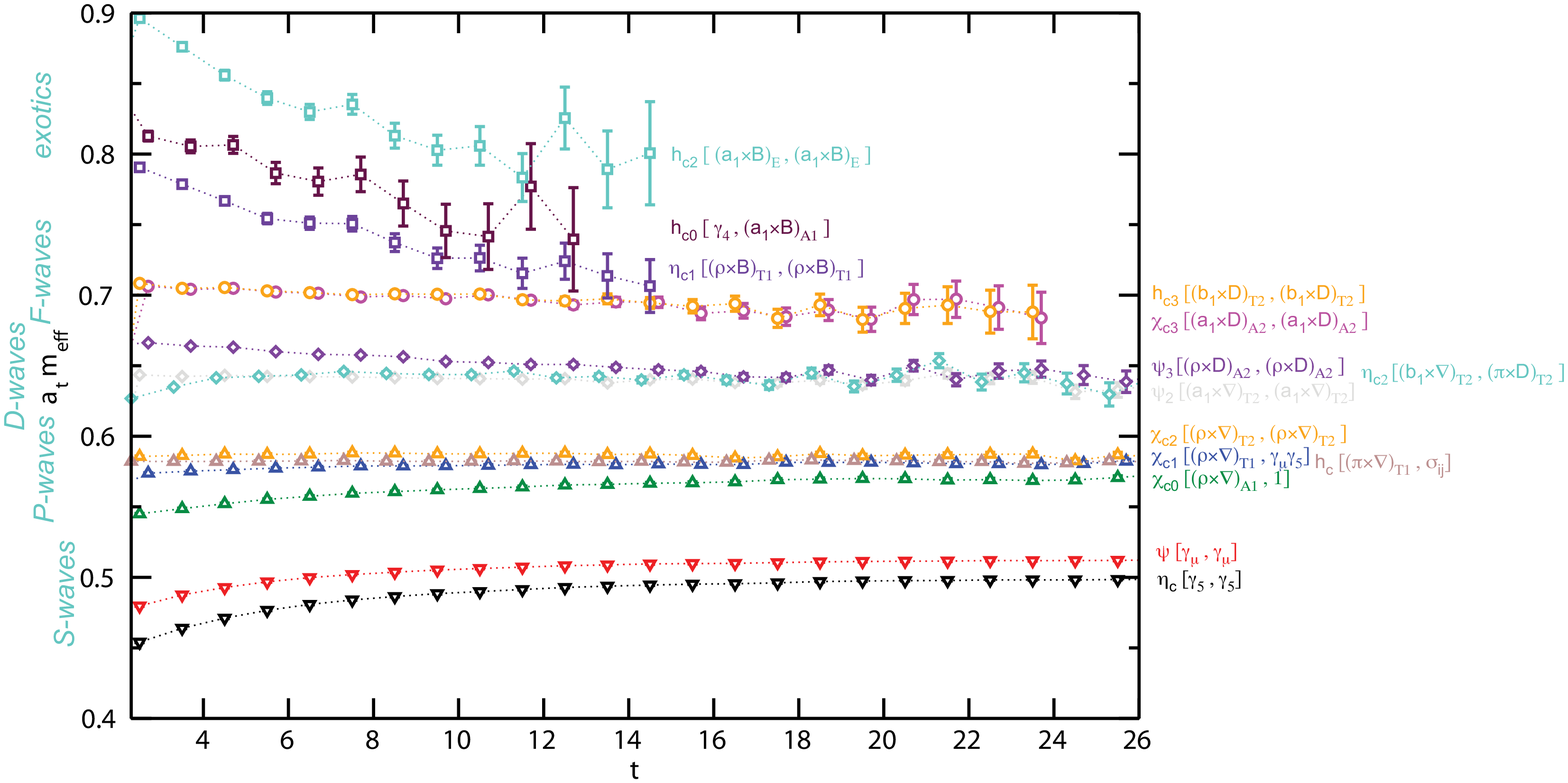}
  \end{center}
\vspace*{-0.2in}
  \caption{Effective masses for various channels showing their relative positions.}
    \label{fig:all}
\end{figure}
\section{Conclusions}
\vspace*{-0.1in}
We present here our preliminary results on exotic and exotic state
mesons in the charmonium sector.  We observe excellent signals for
most channels, including exotics, and a comprehensive analysis with
results for ground and excited states is in preparation.  Our
preliminary observation suggests the presence of the $1^{-+}$ exotic
at around $4.2$~GeV, lower than that was observed in Ref.~\cite{manke}.
The methodology developed here will be utilized in the light quark
sector with dynamical clover fermions. Prediction of masses and particularly
photo-couplings in the light quark sector will be quite helpful to the
GlueX experiments at JLab.

\section*{Acknowledgments}
\vspace*{-0.1in}
Computations were performed using the Chroma software
suite~\cite{chroma} on clusters at Jefferson Laboratory using time
awarded under the SciDAC Initiative.  This work was supported by US
DOE contract DE-AC05-06OR23177 under which Jefferson Science
Associates, LLC administers Jefferson Laboratory.


\end{document}